\documentclass[aps,pra,twocolumn,showpacs]{revtex4-1}
\usepackage{amsmath}
\usepackage{amsfonts}
\usepackage{amssymb}
\usepackage{cancel}
\usepackage{verbatim}
\usepackage{hyperref}
\usepackage{blkarray}
\usepackage{hyperref}
\usepackage{color}

\newcommand{\cA}{{\mathcal A}}
\newcommand{\cB}{{\mathcal B}}
\newcommand{\cC}{{\mathcal C}}
\newcommand{\cK}{{\mathcal K}}
\newcommand{\cL}{{\mathcal L}}
\newcommand{\cR}{{\mathcal R}}

\newcommand{\cH}{{\mathcal H}}
\newcommand{\cN}{{\mathcal N}}
\newcommand{\cM}{{\mathcal M}}
\newcommand{\cT}{{\mathcal T}}

\newcommand{\cQ}{{\mathcal Q}}
\newcommand{\Tr}[1]{{\textrm{Tr}}{\left\{#1\right\}}}
\newcommand{\ket}[1]{|{#1}\rangle}

\newcommand{\ketbra}[2]{|{#1}\rangle\langle{#2}|}
\newcommand{\braket}[2]{\langle{#1}|{#2}\rangle}

\begin{document}
\title{Closed form solution of Lindblad master equations without gain
}
\author{Juan Mauricio Torres}
\email{mauricio.torres@physik.tu-darmstadt.de}
\affiliation{Theoretische Physik, Universit\"at des Saarlandes, 66123 Saarbr\"ucken, Germany}
\affiliation{
Institut f\"ur Angewandte Physik, Technische Universit\"at Darmstadt, 
64289 Germany
}
\affiliation{Departamento de Investigaci\'on en F\'isica, Universidad de Sonora, Hermosillo, M\'exico}
\date{\today}
\begin{abstract}
We present a closed form solution to the eigenvalue problem of
a class of master equations that describe open quantum systems with loss
and dephasing but without gain.
The method relies on the existence
of a conserved number of excitation in the Hamiltonian part
and that none of the Lindblad operators describe an excitation 
of the system. 
In the absence of dephasing Lindblad operators, the eigensystem
 of the Liouville operator can be constructed from the eigenvalues
and eigenvectors of the effective non-Hermitian Hamiltonian used in
the quantum jump approach. 
Open versions of spin chains, 
the Tavis-Cummings model and coupled Harmonic oscillators 
without gain can be solved using this technique.
\end{abstract}


\maketitle
\section{Introduction}
Master equations in Lindblad form provide the most general 
dynamical description of open quantum systems under the Markov assumption
\cite{Lindblad1976,Carmichael1993,Gardiner2004}.  
Sometimes  also called Kossakowski-Lindblad equations, due to
pioneering works of Kossakowski \cite{Kossakowski1972, Gorini1976}, these
type of equations have  been extensively used to describe atom cooling
\cite{Cirac1992}, decoherence in quantum information
theory and quantum engineering of states\cite{Kraus,Verstraete2009}. 

Many Hamiltonian systems have been extended to include dissipation 
in Lindblad form, however, even if the Hamiltonian  part of the system is solvable, 
the full solution to the dissipative version is not obvious. 
The problem results from the fact that the Liouville operator
governing the dynamics is a non-Hermitian operator which
acts on density matrices.
Efforts have been made to tackle this problem
and,  
while there exist analytical  steady state solutions to some problems \cite{Prosen2011},
there are not many  solutions to the eigenvalue problem of specific systems.
The dissipative version of the Harmonic oscillator, up to two  spins 
and the Jaynes-Cummings model
are the only open systems for which exact solutions of the eigenvalue problem are known
\cite{Briegel1993,Barnett2000,Daeubler1992,Kuang1997,VanWonderen1997,Englert2002}.
The solutions to the Jaynes-Cummings model that can be found in the literature
\cite{Briegel1993,VanWonderen1997} are examples that
show how intricate the calculation of the eigensystem of the Liouville operator can be.
In particular, the work of Briegel and Englert \cite{Briegel1993} solves
this problem in terms of the eigenbases of the uncoupled subsystems
which are formed by a damped harmonic oscillator and a damped two level atom.
The interaction part operates in a non trivial way on the elements of these combined bases
and therefore, although  manageable in this case, this procedure is not 
suitable to generalize to higher dimensional systems as it would lead to very 
tedious calculations.

In this work we present a systematic method for solving the eigenvalue
problem of a broad class of Lindblad master equations which do not involve 
any form of gain and that share the characteristic of being solvable in the 
Hamiltonian part with an additional constant of motion that
measures the number of excitations in the system.
Under these assumptions we are able to find the eigenvalues and the eigenbasis
of the operator that is obtained by subtracting
from the  Liouville operator, the jump operator of the quantum jumps approach
\cite{Plenio1998,Carmichael1993}.
We use an expansion in terms of  the elements of this basis to solve for the eigensystem of
the complete Liouville operator and obtain a first order vector recurrence
relation that  can be solved in an iterative way. 
In the absence of dephasing Lindblad operators, the
eigensystem of the full Liouvillian can be constructed in a systematic way
from the eigensystem of the non Hermitian Hamiltonian of the quantum jumps approach. 
Specifically, it is shown that each eigenvalue of the complete system is 
proportional to the sum of two eigenvalues of the corresponding non Hermitian Hamiltonian.
The main difference with respect to the procedure used in Ref. \cite{Briegel1993} is that
we use the eigenbasis of the part of the master equation without the jump operator, 
instead of the eigenbasis without the interaction term.
With this approach we are able to reproduce previous specific solutions to the damped harmonic 
oscillator  \cite{Barnett2000}
and the damped Jaynes-Cummings model \cite{Briegel1993}, but most importantly 
our method is presented in a general way that encompasses 
systems such as the dissipative version of: Heisenberg $XXZ$ spin chains
\cite{Klumper1993}, AKLT model \cite{Affleck1987}, 
Bose-Hubbard model \cite{Kraus},
Tavis-Cummings model \cite{Tavis1968}, etc. 
Our construction is performed for systems that do not present any source of gain, 
nevertheless, similar arguments lead to exact solutions for analogue systems without loss.

The paper is organized as follows. In Sec. \ref{main} we formalize the assumptions that
define the class of systems we want to address. Furthermore, we show the procedure
to solve the eigensystem of the master equation in terms of the eigensystem of an effective 
non-Hermitian Hamiltonian. 
In Sec. \ref{dephasingsection} we consider systems that include depahasing 
Lindblad operators and explain how to solve this class of systems in connection
to the method presented in Sec. \ref{main}. Finally, in Sec. \ref{examples} we present two examples
of systems that can be solved using this technique:
The Jaynes-Cummings model and the two-atoms Tavis-Cummings model.
As an application we evaluate the atomic spontaneous emission spectrum of the first example.

\section{The master equation}
\label{main}
We consider systems whose dynamics is 
governed  by a master equation, which consist of
a coherent Hamiltonian evolution and a dissipative
part in Lindblad form. The dynamical equation is given in terms of the so-called
Liouville operator as   
\begin{equation}\label{master0}
  \hspace{-3pt}\cL\rho
  =\frac{1}{i\hbar}\left[H,\rho\right]
  +\hspace{-3pt}\sum_{s}\frac{\gamma_s}{2}
  \left(2A_s\rho A_s^\dagger-A_s^\dagger A_s\rho-\rho A_s^\dagger A_s\right).
\end{equation}
Concerning the Hamiltonian part we assume that no source of driving 
is present in the system and that an observable 
$I$ exists which commutes with $H$. This additional 
constant of motion may be interpreted as
a measure of the excitations in the system. 
Furthermore, we consider that the system is defined on a Hilbert space $\cH$
and that there exists a complete basis set $\{\ket{n,j}\}$ with $\cN$ elements.
In this basis $I$ is diagonal and there are $d_n$ states with the same 
integer eigenvalue of $I$, 
that is $I\ket{n,j}=n\ket{n,j}$, with $n=0,\dots N$ and $\cN=\sum_{n=0}^N d_n$. 
We make all the treatment for finite $N$, but 
under the same line of thought of references \cite{Briegel1993,Barnett2000,Englert2002},
the results also apply in the limit
$N\to\infty$, which also implies the limit of $\cN\to\infty$, 
for countable infinite separable Hilbert spaces. 
As an immediate consequence of the existence of the conserved quantity 
 $I$ we can identify that the Hamiltonian has a block diagonal form
in the basis where $I$ is diagonal, with each block $H^{(n)}$ of size $d_n\times d_n$.
This feature is essential and will be exploited in our construction.

For the dissipative part, we first consider only Lindblad operators $A_s$ which describe losses
in the system. 
We formalize this  condition
with the commutation relation 
\begin{align}
[A_s,I]=A_s.
  \label{comA}
\end{align}
If we were considering the Jaynes-Cummings model \cite{Jaynes1963}, we 
could take the electromagnetic mode annihilation operator $a$
and the spin lowering operator $\sigma^-$ as Lindblad operators, 
corresponding to the situation when the system interacts with a 
zero temperature reservoir. In this example, the additional
constant of motion would be $a^\dagger a+\sigma^+\sigma^-$.

From the previous consideration it follows that each Hermitian operator  $A_s^\dagger A_s$ also
commutes with $I$, which implies a block diagonal form in the
basis where $I$ is diagonal. This motivates rewriting the master 
equation in two parts, one that conserves the excitations and the
other describing the de-excitation of the system. With the introduction of 
the following non-Hermitian Hamiltonian
\begin{equation}
  K=H
  -i\hbar\sum_s\frac{\gamma_s}{2}A_s^\dagger A_s,
  \label{effHam}
\end{equation}
which can be recognized as the effective Hamiltonian used in the
quantum trajectories technique \cite{Carmichael1993,Plenio1998},
one can rewrite the equation \eqref{master0} as the sum of the
following two parts
$\cL \rho=\cK\rho+\cA\rho$, with
\begin{align}
&  \cK\rho=
  \frac{1}{i\hbar}\left(
  K\rho-\rho K^\dagger
  \right),\quad
  \cA\rho=
  \sum_s \gamma_s A_s\rho A_s^\dagger.
  \label{master1}
\end{align}
The first term describes the part of the dynamics that conserves excitation, while
the second one, the jump operator, describes the de-excitations in the system.

In order to solve the master equation, our strategy will be to find
the eigensystem of $\cK$. Then,
we will deduce how the jump operator acts on each of its eigenvectors. 
Making a plausible Ansatz for the eigenvectors of the complete master equation as a 
superposition of the eigenvectors of $\cK$ and then inserting
them into the full master equation, will allow us to find a solvable 
recursion relation for the coefficients of the superposition.

\subsection{Eigensystem of $K$}
Given the fact that  $[ K,I]=0$  it follows that 
$ K \ket{n,k}=\sum_{j=1}^{d_n} K_{j,k}^{(n)} 
\ket{n,j}$,  i.e. it does not couple
eigenvectors of $I$ with different values of $n$. In the basis
$\{\ket{n,j}\}$, $ K$ has a block diagonal form, with 
each block given by a matrix of size $d_n\times d_n$. 
We assume that each block can be diagonalized by the transformation
\begin{align}\label{vtrans}
  \tilde K^{(n)}=Q^{\dagger(n)} K^{(n)} R^{(n)}, \quad {\rm with}\quad  
  Q^{\dagger(n)}R^{(n)}=\mathbb{I}_{d_n},
\end{align}
where $\tilde K^{(n)}$ is a diagonal matrix with the eigenvalues
of the $n$-th block in its diagonal and $\mathbb{I}_{d_n}$ is the identity
matrix of dimension $d_n$.
We use  a tilde throughout  this manuscript
to denote when a matrix is expressed in the eigenbasis of $K$.
The $d_n\times d_n$ matrices $Q^{\dagger(n)}$ and $R^{(n)}$ are the blocks
of the operators $Q^\dagger$ and $R$ which diagonalize
the operator $K$.
The columns (rows) of $R$ ($Q^\dagger$) are the right (left) 
eigenvectors of $K$ \cite{Note1}
and in the original basis $\{\ket{n,j}\}$, they can be expanded as 
\begin{align}
  \ket{r_j^n}
  &=\sum_{k=1}^{d_n}
  R_{k,j}^{(n)}\ket{n,k},
  &
  \ket{q_j^n}
  &=\sum_{k=1}^{d_n}
  Q_{k,j}^{(n)}\ket{n,k}.
  \label{evecK}
\end{align}
One can verify that these states are also eigenstates of $I$, i.e.
$I\ket{r_j^n}=n\ket{r_j^n}$, and 
assuming that the transformation in  equation
\eqref{vtrans} exists, it follows that they are orthogonal and complete
\begin{align}
  \braket{q_k^n}{r_j^m}=\delta_{k,j}\delta_{n,m},
  \quad 
  \sum_{n=0}^N\sum_{j=1}^{d_n}\ketbra{r_j^n}{q_j^n}
  =\mathbb{I}.
  \label{comrel}
\end{align}
The eigenvalue equation for $ K$ is then
\begin{align}
   K\ket{r_j^n}=\varepsilon_j^{(n)}\ket{r_j^n},
   \quad
   K^\dagger\ket{q_j^n}=
\varepsilon_j^{\ast(n)}
  \ket{q_j^n},
  \label{eigsH}
\end{align}
with the complex eigenvalues $\varepsilon^{(n)}_j$.

\subsection{Eigensystem of $\cK$}
The eigensystem of the operator $\cK$ can be constructed from the eigensystem
of the non-Hermitian Hamiltonian $K$. 
It can be verified by inspection of Eq. \eqref{master1} that the elements
\begin{align}
  \hat\varrho_{j,k}^{(l,n)}=  
  \ketbra{r_{j}^{n+l}}{r_{k}^{n}},
\quad
  \check\varrho_{j,k}^{(l,n)}=  
\ketbra{q_{j}^{n+l}}{q_{k}^{n}},
  \label{eivLH}
\end{align}
with $j=1,\dots d_{n+l}$ and $k=1,\dots d_{n}$,
are the right and left eigenvectors of $\cK$ and that they 
solve the eigenvalue equation 
\begin{align}
 \cK
  \hat\varrho_{j,k}^{(l,n)}
  =
  \lambda_{j,k}^{(l,n)}
  \hat\varrho_{j,k}^{(l,n)},
\qquad
  \cK^\dagger
  \check\varrho_{j,k}^{(l,n)}
  =
  \lambda_{j,k}^{\ast(l,n)}
  \check\varrho_{j,k}^{(l,n)},
  \label{eieqK}
\end{align}
with eigenvalues 
\begin{equation}\label{eigv}
  \lambda_{j,k}^{(l,n)}
  =
  \frac{1}{i\hbar}\left[
  \varepsilon_j^{(n+l)}-
  \varepsilon_{k}^{\ast(n)}
  \right].
\end{equation}
The dual operator of $\cK$ is given by
$
  \cK^\dagger
  \rho 
  =\frac{1}{i\hbar}
  \left(
  \rho K- K^\dagger \rho
  \right)
  $ \cite{Briegel1993,Barnett2000}. 
From equation \eqref{comrel}, it follows that
these eigenvectors  are orthogonal 
with respect to the Hilbert-Schmidt inner product
\begin{align}
  \Tr{
  \left(\check\varrho_{j,k}^{(l,n)}\right)^\dagger
  \hat\varrho_{j',k'}^{(l',n')}
  }=
  \delta_{n,n'}\delta_{l,l'}\delta_{j,j'}\delta_{k,k'}.
  \label{}
\end{align}

Now, let us study more in detail the operator $\cK$ and how it acts
acts on the elements $\ketbra{n+l,j}{n,k}$ with
$n+l,n=0\dots N$, $j=1\dots d_{n+l}$ and $k=1,\dots d_n$.
These elements form a basis
for the vector space $\cB(\cH)$ of the operators that act
on the Hilbert space $\cH$. As $K$ does not couple basis elements of different $n$ it follows
that $\cK$ does not couple elements  with different pairs of excitation numbers
$n+l$ and $n$, that is
\begin{align}
\cK
\ketbra{n+l,j'}{n,k'}
=\sum_{j,k=1}^{d_{n+l},d_n}
\cK^{(l,n)}_{j,k,j',k'}
\ketbra{n+l,j}{n,k}.
\label{blockcK}
\end{align}
This shows that the operator $\cK$ is formed by the uncoupled blocks
$\cK^{(l,n)}$, where each one of them can be represented by 
a tensor of rank $4$ and dimensions 
$d_{n+l}\times d_n\times d_{n+l}\times d_n$. 

To simplify the evaluation we will adopt the following bijective mapping of indices 
$j,k\to \nu$, with
\begin{align}
  \nu=d_n(j-1)+k, \quad \nu=1\dots D_{l,n}=d_{n+l}d_n.
  \label{mapp}
\end{align}
In this convention that maps  two indices to one, 
the tensor in Eq. \eqref{blockcK} can now be expressed as a $D_{l,n}\times D_{l,n}$ matrix
that acts on vectors of size $D_{l,n}$ which are obtained by vectorizing row by row
a matrix of size $d_{n+l}\times d_n$ using the mapping of indices in in Eq. \eqref{mapp}.
With this convention and using the properties of the tensor product,
we can express the blocks of $\cK$ as
\begin{align}
  \cK^{(l,n)}=\frac{1}{i\hbar}
  \left(
  K^{(n+l)}
  \otimes \mathbb{I}_{d_n} 
  -
  \mathbb{I}_{d_{n+l}}
  \otimes 
  K^{\ast(n)}\right).
  \label{outercK}
\end{align}
Analogous to Eq. \eqref{vtrans}, there exists a transformation
which diagonalizes each block of $\cK$. 
It has the form
\begin{align}
  \tilde \cK^{(l,n)}=\cQ^{\dagger(l,n)}\cK^{(l,n)}\cR^{(l,n)},
  \,\,\,\, 
  \cQ^{\dagger(l,n)}
  \cR^{(l,n)}=\mathbb{I}_{D_{l,n}}.
  \label{vstrans}
\end{align}
The eigenvectors of $\cK$, given in the
 Eq. \eqref{eivLH},  provide us with the transformation 
 that diagonalizes each of its blocks $\cK^{(l,n)}$, 
it is given by the tensor product of the matrices with the eigenvectors of $K$, 
i.e.
\begin{align}
  &\cQ^{(l,n)}=Q^{(n+l)}\otimes Q^{\ast(n)},\nonumber\\
  &\cR^{(l,n)}=R^{(n+l)}\otimes R^{\ast(n)}.
  \label{bigbasis}
\end{align}

\subsection{Jump operator}
\label{jumpsection}
We proceed to study the action of the jump operator $\cA$ on the
eigenvectors of $\cK$. As it is formed by the Lindblad operators
$A_s$ we first focus on how these 
act on the eigenvectors of the non-Hermitian Hamiltonian $K$.
We assume that the action of each $A_s$ on states in 
the original basis is known. Considering
its commutation relation with $I$ given in Eq. \eqref{comA} we can deduce that 
it is of the form $A_s\ket{n,j}=\sum_{k=1}^{d_{n-1}} A_{s;k,j}^{(n)}\ket{n-1,k}$.
It is manifested in this way that every $A_s$  connects states of the block
$n$ to states in the block $n-1$, meaning that the Lindblad operators
are also composed of uncoupled blocks $A^{(n)}_s$ of dimension $d_{n-1}\times d_n$.
Using the transformations of  Eq. \eqref{vtrans} it is possible to transform these 
blocks to the representation in the eigenbasis of $K$ in the following way
\begin{align}
\tilde A_{s}^{(n)}=Q^{\dagger(n-1)}A_s^{(n)}R^{(n)}. 
  \label{transA}
\end{align}
Thereby, we find that the action of the Lindblad operators onto the right eigenstates of $K$ can
be expressed as
\begin{align}
  A_s\ket{r^n_j}=
  \sum_{k=1}^{d_{n-1}}\tilde A_{s;k,j}^{(n)}\ket{r_k^{n-1}}.
  \label{actA}
\end{align}

With the blocks of the Lindblad operators in the representation of the eigenbasis of $K$
it is now possible to build the   blocks of the jump operator  using the tensor product.
They have the form
\begin{align}
  \tilde\cA^{(l,n)}=\sum_s \gamma_s \tilde A_s^{(n+l)}\otimes \tilde A_s^{\ast(n)}
  \label{tensA}
\end{align}
and they are matrices of size $D_{l,n-1}\times D_{l,n}$ that connect
vectorized matrices of dimension $D_{l,n}$
to others of dimension $D_{l,n-1}$. 
It is in this representation that one can identify 
how the jump operator acts on the eigenbasis of $\cK$, that is
\begin{align}
  \cA
  \hat\varrho_{\nu}^{(l,n)}
  &=
  \sum_{\nu'}^{D_{l,n-1}}
  \tilde\cA_{\nu',\nu}^{(l,n)}
  \hat\varrho_{\nu'}^{(l,n-1)}.
  \label{jumpr}
\end{align}
Analogously we can find the corresponding equation for the dual jump operator  
acting on the left eigenvectors as
\begin{align}
  \cA^\dagger\check\varrho_{\nu}^{(l,n)}
  &=
  \sum_{\nu'}^{D_{l,n+1}}\tilde\cA_{\nu,\nu'}^{\ast(l,n+1)}
  \check\varrho_{\nu'}^{(l,n+1)}.
  \label{jumpl}
\end{align}
We have adopted the mapping of indices in equation \eqref{mapp} to label
the eigenvectors of $\cK$ and we introduced the dual of the jump operator, defined 
as $\cA^\dagger\rho=\sum_s\gamma_s A_s^\dagger \rho A_s$. 

In an alternative calculation one could start with the evaluation 
of the blocks $\cA^{(l,n)}$ 
in the original basis based 
on the blocks $A^{(n)}_s$, in the same manner as in Eq. \eqref{tensA}. Then 
one could change the basis using the transformation of Eq.
\eqref{bigbasis} to find 
\begin{align}
  \tilde\cA^{(l,n)}=\cQ^{\dagger(l,n-1)}\cA^{(l,n)}\cR^{(l,n)}.
  \label{transcA}
\end{align}

\subsection{Eigensystem of the full master equation}
\label{fullsolution}
Noting that the jump operator  \eqref{jumpr} 
couples eigenvectors of $\cK$ of definite excitation number $n$ with a superposition of
eigenvectors of $n-1$ without changing the value of $l$, it seems reasonable to take as an
Ansatz for the eigenvectors of the full Liouvillian $\cL$ a 
superposition of eigenvectors of $\cK$ with a fixed value of $l$. 
The proposed  Ansatz, in the vectorized convention is 
$\hat\rho^{l,\Lambda}=\sum_{n,\nu} \tilde v^{l,\Lambda;n}_{\nu}
\hat\varrho_{\nu}^{(l,n)}$, where $\Lambda$ is an  eigenvalue
of the full master equation and for the moment it labels the eigenvectors and
its coefficients.
Our next step is to study how the full Liouvillian $\cL$ acts on  these type of 
states. From equations \eqref{jumpr}, \eqref{eieqK} and \eqref{mapp} one obtains
\begin{align}
  \cL\hat\rho^{l,\Lambda}&=\Lambda\hat\rho^{l,\Lambda}
  =
  \sum_{n=0}^N
  \sum_{\nu=1}^{D_{l,n}}
  \tilde v_{\nu}^{l,\Lambda;n}
  \lambda_{\nu}^{(l,n)}
  \hat\varrho_{\nu}^{(l,n)}
  \nonumber\\
  &+
  \sum_{n=1}^{N}
  \sum_{\nu,\nu'=1}^{D_{l,n},D_{l,n-1}}
  \tilde v_{\nu}^{l,\Lambda;n}
  \tilde\cA_{\nu',\nu}^{(l,n)}
  \hat\varrho_{\nu'}^{(l,n-1)}.
\end{align}
Reordering of indices and matching the elements $\hat\varrho_{\nu}^{(l,n)}$ leaves us with the following recurrence relation for the coefficients  at fixed $n$
\begin{align}
\left(\Lambda -
  \lambda_{\nu}^{(l,n)}
  \right)
  \tilde v_{\nu}^{l,\Lambda;n}
  =
  \sum_{\nu'}^{D_{l,n+1}}
  \tilde\cA_{\nu,\nu'}^{(l,n+1)}\tilde v_{\nu'}^{l,\Lambda;n+1}.
  \label{coefrecr}
\end{align}
The relation holds for any complex value of $\Lambda$, but
we take the simplest one in which the recurrence ends, 
using a similar reasoning as in \cite{Barnett2000}. 
We find that the
eigenvalues for the complete Liouville operator are
$\Lambda=\lambda_{\mu}^{(l,m)}$, for certain $m$ and $\mu$ which label
inner blocks in the same way as $\nu$.
This result tells us  that  $\cL$ and $\cK$ have the same spectrum, a fact
that can also be understood as $\cL$ has an upper triangular form in the
basis where $\cK$ is diagonal. 
Another observation is that for $n=m$ the left hand side of 
equation \eqref{coefrecr} vanishes, which means that all coefficients
are zero for $n>m$. The first non vanishing coefficient is  
$\tilde v_{\mu}^{l,\Lambda;m}=1$. 
From here one can proceed to evaluate the rest of the
coefficients in a recursive way.
Note also that $3$ integers are needed to define
each eigenvector: $m$, $\mu$ and $l$ 
(or $4$ if instead one uses $\mu\to (j''-1)d_m+k''$ in the matrix representation). 
Hence, we redefine the coefficients as 
$\tilde v^{l,\Lambda;n}_{\nu} \to \tilde v_{\mu;\nu}^{(l,m;n)}$.

The recursion relation can also be cast in terms of matrix multiplication,
if one takes a vector of coefficients $\tilde v_{\mu}^{(l,m;n)}$ for each
block of $n$. 
Let us define the non-zero elements of the $D_{l,n}\times D_{l,n}$ diagonal
matrix as
\begin{align}
  \tilde\cT_{\mu;\nu,\nu}^{(l,m; n)}=
  (\lambda^{(l,m)}_{\mu}-\lambda^{(l,n)}_\nu)^{-1}.
  \label{diag}
\end{align}
With this definition, 
the recursion in \eqref{coefrecr}  can be solved to give the $n$-th vector with 
$D_{l,n}$ entries 
\begin{align}
  \tilde v_{\mu}^{(l,m;n)}=
  \left(
  \prod_{i=n}^{m-1}
  \tilde\cT^{(l,m;i)}_{\mu}
  \tilde\cA^{(l,i+1)}  
  \right)
  e_\mu^{(l,m)}.
  \label{coefv}
\end{align}
Thereby $\tilde v_{\mu}^{(l,m;m)}=e_\mu^{(l,m)}$ is a column vector 
of dimensions $D_{l,m}$ with a $1$ in the $\mu$-th entry and zero elsewhere.
All the coefficients for $n>m$ vanish.
Now one can write the right eigenvectors of the full Liouvillian as
\begin{align}
  \hat\rho^{(l,m)}_{\mu}
  =\sum_{n=0}^m
  \sum_{\nu=1}^{D_{l,n}} 
  \tilde v_{\mu;\nu}^{(l,m;n)}\hat\varrho_{\nu}^{(l,n)}.
  \label{}
\end{align}

The left eigenvectors can be evaluated in a similar way and
as we already know the eigenvalues of $\cL$ we can 
use a superposition of the left eigenvectors
of $\cK$ with fixed $l$ to find
\begin{align}
  \cL^\dagger\check\rho^{(l,m)}_\mu&=
  \lambda^{\ast(l,m)}_\mu\check\rho^{(l,m)}_\mu
  =
  \sum_{n=0}^N
  \sum_{\nu=1}^{D_{l,n}}
  \tilde u_{\mu;\nu}^{(l,m;n)}
  \lambda_{\nu}^{\ast(l,n)}
  \check\varrho_{\nu}^{(l,n)}
  \nonumber\\
  &+
  \sum_{n=0}^{N-1}
  \sum_{\nu,\nu'=1}^{D_{l,n},D_{l,n+1}}
  \tilde u_{\mu;\nu}^{(l,m;n)}
  \tilde\cA_{\nu,\nu'}^{\ast(l,n+1)}
  \check\varrho_{\nu'}^{(l,n+1)}.
\end{align}
Again, reordering indices and matching
the coefficients for each $\check\varrho_\nu^{(l,m)}$ 
we find the recursion relation 
\begin{align}
  \left(\lambda^{\ast(l,m)}_\mu - \lambda^{\ast(l,n)}_\nu
  \right)
  \tilde u
  ^{(l,m;n)}
  _{\mu;\nu}
  =
  \sum_{\nu'=1}^{D_{l,n-1}}
  \tilde\cA
  ^{\ast(l,n)}
  _{\nu',\nu}
  \tilde u
  ^{(l,m;n-1)}_{\mu;\nu'},
  \label{coefrecl}
\end{align}
which can be iterated  to give the solution for the coefficients as
\begin{align}
  \tilde u_{\mu}^{(l,m;n)}=
  \left(
  \prod_{i=n-1}^{m}
  \tilde\cT^{\dagger(l,m;i+1)}_{\mu}
  \tilde\cA^{\dagger(l,i+1)}
  \right)
   e_\mu^{(l,m)}.
   \label{coefu}
\end{align}
In this way, we find the following expression for the left eigenvectors
\begin{align}
  \check\rho^{(l,m)}_{\mu}
  =\sum_{n=m}^N
  \sum_{\nu=1}^{D_{l,n}} 
  \tilde u_{\mu;\nu}^{(l,m;n)}\check\varrho_{\nu}^{(l,n)}.
  \label{}
\end{align}

To express the eigenvectors of the full Liouvillian in the original basis,
one can apply the transformation in Eq. \eqref{bigbasis} one by one to 
each of the vectors in equations 
\eqref{coefv} and \eqref{coefu} as 
\begin{align}
  v_\mu^{(l,m;n)}=
  \cR^{(l,n)}\tilde 
  v_\mu^{(l,m;n)},
  \,\,\,
  u_\mu^{(l,m;n)}=
  \cQ^{(l,n)}\tilde 
  u_\mu^{(l,m;n)}.
  \label{}
\end{align}
Using the mapping of indices in equation \eqref{mapp}, 
one finally finds the left and right set of eigenvectors
in the original basis,
\begin{align}
  \hat\rho^{(l,m)}_{j,k}
  =\sum_{n=0}^m
  \sum_{j',k'=1}^{d_{n+l},d_n} 
  v_{j,k;j',k'}^{(l,m;n)}
  \ketbra{n+l,j'}{n,k'},
  \nonumber\\
  \check\rho^{(l,m)}_{j,k}
  =\sum_{n=m}^N
  \sum_{j',k'=1}^{d_{n+l},d_n} 
  u_{j,k;j',k'}^{(l,m;n)}
  \ketbra{n+l,j'}{n,k'}.
  \label{finaleig}
\end{align}
In a matrix representation the right eigenvectors take the following form
\begin{align*}
  \hat\rho_{j,k}^{(l,m)}=\,\,
  \begin{blockarray}{c|c|c|c|ccc}
    0 & \dots&m&m+1&\dots \\
    &&&\\
    \begin{block}{(c|c|c|c|c)cc}
      0 & \dots & &&&0 \\
      \cline{1-7}
      \vdots &\ddots&  & &&\vdots \\      
      \cline{1-7}
      v^{(l,m;0)}_{j,k} &  & && & l \\
      \cline{1-7}
       &\ddots&   &&  & \vdots\\
      \cline{1-7}
       & &v^{(l,m;m)}_{j,k}  & &&   m+l \\
      \cline{1-7}
       &&& 0 & &  \,\,m+l+1 \\
      \cline{1-7}
       && & & \ddots  &   \vdots\\
    \end{block}
  \end{blockarray}
\end{align*}
where it is manifested that they are formed by uncoupled blocks that lie 
in the $l$-th diagonal
and there are non-zero entries only until the excitation value $m$.
The left eigenvectors can be represented as
\begin{align*}
  \check\rho_{j,k}^{(l,m)}=\,\,
  \begin{blockarray}{c|c|c|c|ccc}
    0  &\dots&m&m+1&\dots \\
    &&&\\
    \begin{block}{(c|c|c|c|c)cc}
      0 & \dots & &&&0 \\
      \cline{1-7}
      \vdots &\ddots  & &&&\vdots \\      
      \cline{1-7}
      0 &  & &&& l+1\\
      \cline{1-7}
       &\ddots& &&  & \vdots\\
      \cline{1-7}
       & &u^{(l,m;m)}_{j,k}  & &&   m+l \\
      \cline{1-7}
       &&  & u^{(l,m;m+1)}_{j,k}  & &\,\, m+l+1 \\
       \cline{1-7}  
       && & & \ddots  & \vdots  \\
    \end{block}
  \end{blockarray}.
\end{align*}
In this case they also lie in the $l$-th diagonal, but in contrast 
to the right eigenvectors,
the non-zero entries start at excitation number $m$.

In this construction we have only focused on elements that lie below the main
diagonal $l=0$. The elements above the
main diagonal can be evaluated from those which have
$l\neq 0$ by taking the 
Hermitian adjoint as
$\hat\rho^{\dagger(l,m)}_{j,k}$ and
$  \check\rho^{\dagger(l,m)}_{j,k}$, with corresponding eigenvalues $\lambda_{j,k}^{\ast(l,m)}$.
Including them 
completes a basis set that spans the  vector space $\cB(\cH)$.

\section{Inclusion of dephasing operators}
\label{dephasingsection}
In this section we briefly comment on the inclusion of dephasing 
Lindblad operators $C_s$ in the master equation. These 
have the commutation relation $[C_s,I]=0$ with the constant of motion $I$.
Its inclusion results in a Liouville operator that we write in the form
$\cL=\cM+\cA$, with $\cM=\cK+\cC$, which is written in terms of the
operators in Eq. \eqref{master1} and the new term 
\begin{align}  
\cC\rho=\sum_{s}\frac{\kappa_{s}}{2}\left(2C_s \rho C_s^\dagger
  -C_s^\dagger C_s\rho -\rho C_s^\dagger C_S\right).
\end{align}
The Liouvillian $\cC$ also preserves the excitation numbers $n$ and $l$,
but
can not be constructed solely out of a non Hermitian Hamiltonian because of the
dephasing term $\sum_s\kappa_s C_s\rho C^\dagger_s$. Nevertheless, 
as we have separated the parts of the Liouvillian $\cL$ that conserves excitations
from the jump operator,
the diagonalization in this case can be carried out in a similar manner as
it was shown in the previous section. In this case the diagonalization of
the blocks of $\cM$ is required. These can be evaluated using Eq. \eqref{outercK} with 
$\cM^{(l,n)}=\cK^{(l,n)}+\cC^{(l,n)}$ and the blocks of $\cC$  which can be constructed as
\begin{align}
 \label{cCblocks}
  \cC^{(l,n)}=&\sum_s\frac{\kappa_s}{2}\left(
 2C_s^{(n+l)}\otimes C_s^{\ast(n)}\right.
 \\&
 \left.
 -\left[C^\dagger_s C_s\right]^{(n+l)}\otimes \mathbb{I}_{d_n}
 -\mathbb{I}_{d_{n+l}}\otimes \left[C^\dagger_s C_s\right]^{\top(n)}
  \right).
 \nonumber
\end{align}
Here we have assumed that the action of each $C_s$ onto the basis $\{\ket{n,j}\}$ is 
known and so the $d_n\times d_n$ matrices $C_s^{(n)}$ are known.
Thereby, the transformations to be found are those that diagonalize
each block as  $\cQ^{\dagger(l,n)}\cM^{(l,n)}\cR^{(l,n)}$ and form the eigenvectors of $\cM$.
From this point, the diagonalization procedure of the full Liouvillian
follows the same steps as in section \ref{fullsolution},
with the eigensystem of $\cK$ being replaced by the eigensystem of $\cM$ and
the blocks of the jump operator evaluated as in Eq. \eqref{transcA}. The
eigenvalues of $\cL$ are the eigenvalues of the excitation-preserving part $\cM$.

\section{Examples}
\label{examples}
In this section we present examples of systems with physical relevance that can be 
solved using the technique introduced in sections \ref{main} and \ref{dephasingsection}.

\subsection{Jaynes-Cummings model}
\label{examplesJC}
The first example we consider is the damped Jaynes-Cummings model
which describes the interaction of a two level system with one
mode of the electromagnetic field inside an optical cavity \cite{Jaynes1963}.
We use this model to test and show our method and compare with the solution
introduced by Briegel and Englert in Ref.
\cite{Briegel1993}.

The Hamiltonian of the JC-model in  the interaction picture with respect to the 
cavity mode energy is given by
\begin{equation}
  H=
  \hbar\delta\sigma^+\sigma^-+ \hbar g(a\sigma^++a^\dagger \sigma^-),
  \label{}
\end{equation}
where $a$ and $a^\dagger$ are the cavity mode creation and annihilation
operators, and $\sigma^\pm=\tfrac{1}{2}(\sigma^x\pm i\sigma^y)$ are 
the raising and lowering operators 
of  the two-level system (TLS) and are defined in terms of the Pauli matrices 
$\sigma^x$, $\sigma^y$ and $\sigma^z$.
The detuning between the TLS frequency gap and the cavity mode 
is given by $\delta$. 

In this case one can recognize that the constant of  motion is given by
\begin{equation}
  I=a^\dagger a+\sigma^+\sigma^-,
  \label{}
\end{equation}
and its eigenbasis $\{\ket{n,j}\}$ is given by the states
\begin{align}
\ket{n,1}&=\ket{n}\otimes\ket{g}, \quad n\ge 0
\nonumber\\
\ket{n,2}&=\ket{n-1}\otimes\ket{e}, \quad n>0, 
\label{basisJC}
\end{align}
with  the number state $\ket{n}$ describing $n$ photons
in the cavity and the atomic excited  $\ket{e}$ and ground state  $\ket{g}$.
 The state $\ket{n,2}$ is only permissible for $n>0$ and so we note that
 the basis is formed by a singlet state, 
which is eigenvector of $I$ with eigenvalue $0$,  
and a family of pairs with eigenvalue $n$. 
Therefore, $d_0=1$ and $d_{n>0}=2$ in this example.

The Lindblad operators we consider are $\sigma^-$ and $a$ which fulfill  the 
commutation relation 
of Eq. \eqref{comA}.
Using them we can construct the effective Hamiltonian of equation \eqref{effHam} which reads
\begin{equation}
  K=H-i\tfrac{1}{2}\hbar\gamma\sigma^+\sigma^--i\tfrac{1}{2}\hbar\kappa a^\dagger a.
  \label{}
\end{equation}
It is a non Hermitian operator which commutes with $I$, so that
one can write it as a block diagonal matrix in the basis of \eqref{basisJC},
with its blocks given by
\small{
\begin{align}
  &K^{(0)}=0, 
  \nonumber\\
  &K^{(n>0)}=\hbar
  \left(
  \begin{array}{cc}
    -i\frac{ n \kappa}{2}
    & g\sqrt{n}
    \\
    g\sqrt{n}&
    \frac{2\delta-i(n-1)\kappa-i\gamma}{2}
  \end{array}
  \right).
\label{blocksJC}
\end{align}
The eigenvalues can be computed and  are given by
\begin{align}
  \varepsilon_j^{(n)}=
  \hbar\tfrac{2\delta-i(2n-1)\kappa-i\gamma}{4}
  +(-1)^j \hbar\sqrt{g^2n+\tfrac{\left(2\delta+i\kappa-i\gamma\right)^2}{16}}.
  \label{eivalJC}
\end{align}
It can be checked that they are degenerate only
in the special case $\delta=0$ and $16g^2 n=(\kappa-\gamma)^2$.
Apart from this case, the eigenvalues
of each block are different and the diagonalization of the matrices $K^{(n)}$ 
can be accomplished with the transformation 
\begin{align}
R^{(n>0)}&=
  \left(
  \begin{array}{cc}
    \cos\theta_n&-\sin\theta_n\\
    \sin\theta_n&\cos\theta_n
  \end{array}
  \right),
  \nonumber\\
\theta_n&=\arctan\left(
\frac{2\varepsilon_1^{(n)}+i n\hbar\kappa}{2\hbar g\sqrt{n}}
\right).
  \label{RJC}
\end{align}
In this example $Q^{\dagger}=R^{\top}$, because $K$ is a complex 
symmetric operator.
Together with Eqs. \eqref{evecK} and \eqref{basisJC} the right and left
eigenvectors of $K$ can be obtained.
For $n=0$ no transformation is needed as 
the singlet $\ket{0,1}$ is already an eigenstate of $K$.

Using Eq. \eqref{outercK} together with \eqref{blocksJC} one can build the 
blocks $\cK^{(l,n)}$ of the generator $\cK$ defined in Eq. \eqref{master1} which
represents the part of the Liouvillian 
$\cL$ which conserves the number of excitations.
These blocks can be diagonalized by the transformation 
$\cR^{(l,n)}$ which can be obtained from
Eq. \eqref{bigbasis} and \eqref{RJC}. It is again an orthonormal transformation in 
this example. The eigenvalues of $\cK$ are also the eigenvalues of the full
Liouvillian $\cL$ and can be evaluated from the Eq. \eqref{eigv}, 
by inserting the result of  Eq. \eqref{eivalJC}.

The Lindblad operators can also be expressed in terms of blocks
in the basis of Eq. \eqref{basisJC}. They have the following form
\begin{align}
  &{\sigma^-}^{(1)}=
    \left(\begin{array}{cc}
     0&1
   \end{array}\right),
   &{\sigma^-}^{(n>1)}=
    \left(\begin{array}{cc}
      0&1\\
      0&0
    \end{array}\right),
  \nonumber\\
  &a^{(1)}=
    \left(\begin{array}{cc}
     1&0
   \end{array}\right),
  &a^{(n>1)}=
    \left(\begin{array}{cc}
      \sqrt{n}&0\\
      0&\sqrt{n-1}
    \end{array}\right).
  \label{blocksAJC}
\end{align}
Using them as explained in Sec. \ref{jumpsection} one can also construct
the blocks of the jump operator $\cA$. All the information is now assembled
and the full solution can be obtained as explained in Sec. \ref{fullsolution}.

In particular, the part of the eigensystem with lowest excitation number is given by
\begin{align*}
\hat\rho_{1,1}^{(0,0)}&=\ketbra{0,1}{0,1},
&\check\rho_{1,1}^{(0,0)}&=\mathbb{I}, 
\\
\hat\rho_{j,1}^{(1,0)}&=\ketbra{r^1_j}{0,1},
&\check\rho_{j,1}^{(1,0)}&=\ketbra{q^1_j}{0,1}+
\dots
\\
\hat\rho_{j,k}^{(0,1)}&=\ketbra{r_j^1}{r_k^1}-\braket{r^1_k}{r^1_j}\ketbra{0,1}{0,1},
&\check\rho_{j,k}^{(0,1)}&=\ketbra{q_j^1}{q_k^1}+\dots 
\end{align*}
with the corresponding eigenvalues 
$\lambda^{(0,0)}_{1,1}=0$, 
$\lambda^{(1,0)}_{j,1}=-i\varepsilon^{(1)}_j/\hbar$ and
$\lambda^{(0,1)}_{j,k}=-i\left(\varepsilon^{(1)}_j-\varepsilon^{\ast(1)}_k\right)/\hbar$.

We close this subsection by remarking that our method can solve the Jaynes-Cummings model
by solving $2\times 2$ complex matrices which are the blocks of the effective Hamiltonian 
widely used in the quantum trajectories approach. The solutions can be manageable
at least for few excitation numbers as  will be shown in the following application.

\subsubsection{Spontaneous emission spectrum}
As an application of the solution to the damped Jaynes-Cummings model, we
present the evaluation of the spontaneous emission spectrum 
 \cite{Carmichael1989,Auffeves2008}
of the atom which can be written as
\begin{align}
  S(\omega)=\frac{s(\omega)}{2\pi \varsigma}=
  \frac{
  \int_0^\infty dt \int_0^\infty dt' 
  e^{i\omega(t'-t)}\left<\sigma^+(t)\sigma^-(t')\right>
  }
  {2\pi
  \int_0^\infty dt\left<\sigma^+(t)\sigma^-(t)\right>
  },
  \label{spectrumdef}
\end{align}
where we have introduced $\varsigma$, the constant integral which
appears in the denominator and serves as normalization factor. 
To evaluate the correlations functions which appear in Eq. \eqref{spectrumdef},
we can use the relation (see \cite{Carmichael1993})
\begin{align}
  \left<
  \sigma^+(t)\sigma^-(t')
  \right>=\Tr{\sigma^- e^{\cL(t'-t)}\left[\rho(t)\sigma^+\right]},
  \label{correlation}
\end{align}
which can be evaluated with the aid of the eigenbasis of $\cL$ as
one can express the time evolution of any given initial 
state $\rho_0$ in the following form
\begin{align}
  \rho(t)=e^{\cL t}\rho_0=\sum_\lambda
  \Tr{\check\rho_\lambda^\dagger\rho_0}e^{\lambda t}\hat\rho_\lambda.
  \label{rhoevol}
\end{align}
The sum runs over all the eigenvalues 
$\lambda$ which at this stage have not been specified.
Using the results  of Eqs. \eqref{correlation} and \eqref{rhoevol}
together with the definition  of the emission spectrum $S(\omega)$
in Eq. \eqref{spectrumdef} and performing the integration, one obtains
\begin{align}
  s(\omega)=
  \sum_{\lambda,\lambda'}
  \frac{T_{\lambda,\lambda'}}{(\lambda-\lambda'-i\omega)(\lambda'+i\omega)},
  \quad
  \varsigma =
  \sum_{\lambda,\lambda'}\frac{T_{\lambda,\lambda'}}{-\lambda},
  \label{spectrum1}
\end{align}
where we have introduced the weight factors
$  
T_{\lambda,\lambda'}=
  \Tr{\check\rho_\lambda^\dagger\rho_0}
  \Tr{\check\rho_{\lambda'}^\dagger\hat\rho_\lambda\sigma^+}
  \Tr{\sigma^-\hat\rho_{\lambda'}}.
$
If we assume that initially the atom is in the excited state and the cavity
in the vacuum state, that is a state  given by $\rho_0=\ketbra{1,2}{1,2}$,
one can realize that the only contributions to those traces are given by
the eigenvectors with 7 different eigenvalues, namely
$\lambda_{j,k}^{(0,1)}$, $\lambda_{k,1}^{(1,0)}$ and $\lambda_{0,0}^{(0,0)}$
($j,k=1,2$). With this one gets the solution to the emission spectrum as
\begin{align}
  s(\omega)&=
  \left|
  \frac{\hbar\sin^2\theta_1}{\varepsilon^{(1)}_1-\hbar\omega}
  +
  \frac{\hbar\cos^2\theta_1}{\varepsilon^{(1)}_2-\hbar\omega}
  \right|^2
  \nonumber\\
  &=\left|\frac{2(2\omega+i\kappa)}
  {4g^2+(2\delta-2\omega-i\gamma)(2\omega+i\kappa)}\right|^2,
  \label{}
\end{align}
with the normalization factor given by the expression 
\begin{align}
  \varsigma&=
  \frac{\hbar|\sin\theta_1|^4}
  {i(\varepsilon^{(1)}_1-\varepsilon^{\ast(1)}_1)}
  +\frac{\hbar|\cos\theta_1|^4}
  {i(\varepsilon^{(1)}_2-\varepsilon^{\ast(1)}_2)}
  +2{\rm Re}\left[\frac{\hbar\sin^2\theta_1\cos^2\theta_1^\ast}
  {i(\varepsilon^{(1)}_1-\varepsilon^{\ast(1)}_2)}\right]
  \nonumber\\
  &=\frac
  {4g^2(\gamma+\kappa)+\kappa(4\delta^2+(\gamma+\kappa)^2)}
  {4g^2(\gamma+\kappa)^2+\gamma\kappa(4\delta^2+(\gamma+\kappa)^2)}.
  \label{}
\end{align}
This result can be checked to be in agreement with the ones obtained previously
using different methods in Refs. \cite{Carmichael1989,Auffeves2008}.

\subsubsection{Including dephasing Lindblad operators}
The situation is slightly different when dephasing Lindblad operators are included,
those which have the property of commuting with the constant of motion $I$.
In this example we additionally consider  $\sigma^z$ as Lindblad operator, which clearly
has the desired property. To the full Liouvillian one has
 to include the following dissipator 
\begin{align}
  \cC\rho=\gamma_z\left(\sigma^z\rho\sigma^z-\rho\right).
  \label{}
\end{align}
As all the Lindblad operators in $\cC$ commute with $I$, one 
can no longer use the eigensystem of an effective non Hermitian 
Hamiltonian $K$ to construct the
eigenvalues of the full Liouvillian. There is a dephasing 
operator $\sigma^z\rho\sigma^z$
which conserves excitations and whose effect can not be included in $K$.
However, one can exploit the fact that 
$\cC$ conserves the excitations and express it in blocks $\cC^{(l,n)}$.
To this end one needs the representations of $\sigma^z$ in the basis of Eq.
\eqref{basisJC}. One can verify that these 
are ${\sigma^z}^{(0)}=-1$ and 
${\sigma^z}^{(n>0)}=-\sigma^z={\rm diag (-1,1)}$,
a diagonal matrix with entries $-1$ and $1$.
With them one is able to construct the blocks of $\cC$ 
like in Eq. \eqref{cCblocks} as
\begin{align}
  &\cC^{(0,0)}=0,
  \quad \cC^{(l>0,0)}=\gamma_z\left(\sigma^z- \mathbb{I}_2\right),
  \nonumber\\
  &\cC^{(l\ge 0,n>0)}=\gamma_z\sigma^z\otimes\sigma^z-\gamma_z \mathbb{I}_4 .
  \label{}
\end{align}
The operator $\cM=\cK+\cC$ represents the part of the Liouville operator
that conserves excitations and it can be expressed by the uncoupled blocks
$\cM^{(l,n)}=\cK^{(l,n)}+\cC^{(l,n)}$, with each block given by
    \small{
  \begin{align}
    &\cM^{(0,0)}=0,
    \nonumber\\
    &\cM^{(l>0,0)}
    =\frac{1}{i\hbar}K^{(l)} +\gamma_z(\sigma^z-\mathbb{I}_2)   ,
    \nonumber\\
    &\cM^{(l\ge 0,n>0)}=
    -\tfrac{1}{2}\left((2n+l-1)\kappa+\gamma\right)\mathbb{I}_4
    \nonumber\\
    &\,\,\,\,-i\left(
    \begin{array}{cccc}
    i\frac{\gamma-\kappa}{2}&- g \sqrt{n}& g \sqrt{l+n}&0\\
    - g \sqrt{n}&-i 2 \gamma_z - \delta &0& g \sqrt{l+n}\\
     g \sqrt{l+n}&0&-i 2 \gamma_z+ \delta &- g \sqrt{n}\\
    0& g \sqrt{l+n}&- g \sqrt{n}&i\frac{\kappa -\gamma}{2}
  \end{array}\right).
    \label{}
  \end{align}}
It can be verified that the characteristic polynomial of the 
$4\times 4$ blocks shown above, coincide with the one presented in
the work by Briegel and Englert 
for the same situation
\cite{Note2}.
The diagonalization of these blocks $\cM^{(l,n)}$ gives the corresponding
transformation  $\cR^{(l,n)}$, 
which in this case is again orthogonal as the Liouvillian remains
complex symmetric and so $\cQ=\cR^\top$. 
These transformations have to be used together with Eq. \eqref{transcA} and the blocks
of the jump operator defined $\cA^{(l,n)}$ formed with the blocks of
Eq. \eqref{blocksAJC}.
The eigenvalues  $\lambda^{(l,n)}_\nu$ of 
$\cM^{(l,n)}$  are eigenvalues of the full Liouvillian as well.  
With these one can compute the left and right
eigenvectors of the full Liouvillian as indicated in Sec. \ref{fullsolution}.

\subsection{Two atoms Tavis-Cummings model}
As a second example we consider the two atoms Tavis-Cummings model
\cite{Tavis1968} with damping. The model describes two two-level atoms
interacting with one mode of the electromagnetic field of an optical  cavity.
The Hamiltonian that describes this situation can be written as
\begin{align}
  H=\sum_{\ell=1}^2
  \hbar\left(
  \delta_\ell\sigma_{\ell}^{+}\sigma_{\ell}^-
  +g_\ell(
  a\sigma_\ell^{+}+a^\dagger\sigma_\ell^{-}
  )
  \right).
  \label{HTC}
\end{align}
where $\ell$ labels the operators for each of the two atoms, which can be of different species
as we consider possible distinct couplings strengths $g_\ell$  and detunings $\delta_\ell$.
The number of excitations in the system can be described by the constant
of motion
\begin{equation}
  I=a^\dagger a+
  \sigma_1^+\sigma_1^-+
  \sigma_2^+\sigma_2^-.
  \label{TCI}
\end{equation}
The states of the eigenbasis $\{\ket{n,j}\}$ of $I$ can be organized as follows
\begin{align}
\ket{n,1}&=\ket{n}\ket{g}_1\ket{g}_2
\nonumber\\
\ket{n,2}&=\ket{n-1}\ket{g}_1\ket{e}_2,
&\ket{n,3}&=\ket{n-1}\ket{e}_1\ket{g}_2, &n>0
\nonumber\\
\ket{n,4}&=\ket{n-2}\ket{e}_1\ket{e}_2, & n>1.
\label{basisTC}
\end{align}
where one can note that the degeneracy of $I$ can be divided in blocks of 
$d_0=1$, $d_1=3$ and $d_{n>1}=4$ states. 

As Lindblad operators we consider $\sigma_1^-$, $\sigma_2^-$ and $a$ and doing so 
allows us to write the non-Hermitian Hamiltonian as
\begin{align}
K=H
-i\hbar\tfrac{1}{2}\kappa a^\dagger a
-i\hbar\tfrac{1}{2}\sum_{\ell=1}^2\gamma_\ell\sigma_\ell^+\sigma_\ell^-.
\label{KTC}
\end{align}
In the representation of the basis states of Eq. \eqref{basisTC}  
the non-Hermitian Hamiltonian has a block diagonal form with blocks given by
  \small{
\begin{align}
&  K^{(0)}=0,
\quad
  K^{(1)}=
 \hbar\left(
  \begin{array}{ccc}
    -i\tfrac{1}{2}\kappa&g_2&g_1\\
    g_2&\delta_2-i\tfrac{1}{2}\gamma_2&0\\
    g_1&0&\delta_1-i\tfrac{1}{2}\gamma_1
  \end{array}
  \right),
\nonumber\\[.5em]
&  K^{(n>1)}=
-i\hbar\tfrac{1}{2}\left( n\kappa-\kappa+\gamma_1+\gamma_2\right)
\mathbb{I}_4
  \\
 & \quad\quad+\hbar\left(
  \begin{array}{cccc}
    i\frac{\gamma_1+\gamma_2-\kappa}{2}
    &g_2\sqrt{n}&g_1\sqrt{n}&0\\
    g_2\sqrt{n}&\delta_2+i\tfrac{1}{2}\gamma_1&0&g_1\sqrt{n-1}\\
    g_1\sqrt{n}&0&\delta_1+i\tfrac{1}{2}\gamma_2&g_2\sqrt{n-1}\\
    0&g_1\sqrt{n-1}&g_2\sqrt{n-1}&\delta_1+\delta_2+i\tfrac{1}{2}\kappa
  \end{array}
  \right).
  \nonumber
  \label{}
\end{align}}
Because these are at most $4\times 4$ matrices, their eigensystem can
be evaluated in analytical form. The solutions are lengthy and we do not
present them here, but just comment that as this system can be solved
in analytical way, also the full eigensystem of $\cL$ can be evaluated analytically.

The blocks of the Lindblad operators can be expressed as
\begin{align}
  &{\sigma^-_1}^{(1)}=
  \left(\begin{array}{ccc}
  0&0&1
  \end{array}\right),\quad
  {\sigma^-_1}^{(2)}=
  \left(\begin{array}{cccc}
  0&0&1&0\\
  0&0&0&1\\
  0&0&0&0
  \end{array}\right),
  \nonumber\\
  &{\sigma^-_1}^{(n>2)}=
  \left(\begin{array}{cccc}
  0&0&1&0\\
  0&0&0&1\\
  0&0&0&0\\
  0&0&0&0
  \end{array}\right),\quad
  {\sigma^-_2}^{(1)}=
  \left(\begin{array}{ccc}
  0&1&0
  \end{array}\right),
  \nonumber\\
  &{\sigma^-_2}^{(2)}=
  \left(\begin{array}{cccc}
  0&1&0&0\\
  0&0&0&0\\
  0&0&0&1
  \end{array}\right),\quad
  {\sigma^-_2}^{(n>2)}=
  \left(\begin{array}{cccc}
  0&1&0&0\\
  0&0&0&0\\
  0&0&0&1\\
  0&0&0&0
  \end{array}\right)
  \nonumber\\
  \label{}
\end{align}
and for the Lindbald operator $a$ we obtain the following blocks
\begin{align}
&{a}^{(1)}=
  \left(\begin{array}{ccc}
  1&0&0
  \end{array}\right),\quad
  {a}^{(2)}=
  \left(\begin{array}{cccc}
  \sqrt2&0&0&0\\
  0&1&0&0\\
  0&0&1&0
  \end{array}\right),
  \nonumber\\
  &{a}^{(n>2)}=
  \left(\begin{array}{cccc}
  \sqrt n&0&0&0\\
  0&\sqrt{n-1}&0&0\\
  0&0&\sqrt{n-1}&0\\
  0&0&0&\sqrt{n-2}
  \end{array}\right).
  \label{}
\end{align}

This is all the information one needs to evaluate the solution of the eigensystem of
the Liouville operator of this problem following the steps of Sec. \ref{main}. 

We would like to mention that  one could also consider
Heisenberg $XXZ$ interaction between the atoms, as the terms
$\left(  \sigma_1^{+} \sigma_2^{-}+\sigma_1^{-}\sigma_2^{+} \right)$ and 
$\sigma_1^z\sigma_2^z$ also commute with the constant of motion $I$ and the block
structure is preserved \cite{Torres2010}.

\subsection{Other extensions}
In principle one could consider higher dimensional systems that belong to the
class of systems proposed to be solved using the construction explained in this manuscript.
The solution would involve the diagonalization of blocks of larger dimension
and thus not suitable for analytical calculations. Nevertheless this gives a systematic
method of solving the problem efficiently in a numerical way and
gives insight into the behaviour of open systems. Therefore, we briefly comment on
two important class of systems that can be treated in this way.

\subsubsection{$M$ interacting spins}
For a system composed of $M$ interacting spins we could consider
as a constant of motion  $I=\sum_{\ell}^M\sigma_{\ell}^+\sigma_\ell^-$.
In this case the eigenstates of $I$ with eigenvalue $n$ can be chosen as 
all the states with $n$ spins up which  amounts to
$d_n=\binom{M}{n}$ states. This follows from the fact that one has to arrange $n$ spin up 
particles out of a total of $M$.
A possible Hamiltonian in this case could be
$H=\sum_{\ell>\jmath}J_{\ell,\jmath}\sigma_\ell^z\sigma_{\jmath}^z+
\eta_{\ell,\jmath}\left(\sigma_\ell^+\sigma_{\jmath}^-+\sigma_\ell^-\sigma_{\jmath}^+\right)
$. One could also include the  term $(\vec\sigma_\ell\cdot\vec\sigma_\ell')^2$ like in
the AKLT model \cite{Affleck1987}, because such a term also commutes with $I$.
As Lindblad operators one could consider $\{\sigma_\ell^-\}_{\ell=1}^M$ and
the dephasing Lindblad operators $\{\sigma_\ell^z\}_{\ell=1}^M$. 

\subsubsection{$M$ spins interacting with one oscillator}
The situation is similar for $M$ interacting spins with an oscillator, such
as the general Tavis-Cummings model \cite{Tavis1968}.
The constant of motion is in this case
$I=a^\dagger a +\sum_{\ell=1}^M\sigma_{\ell}^+\sigma_\ell^-$.
The degeneracy of the excitation  $n$ can be divided in two classes, one 
with 
$d_{n<M}=\sum_{n'=0}^n \binom{M}{n'}$ and  the other with $d_{n\geq M}=2^M$. This sum
can be understood as for each $n-n'$ excitations in the oscillator there are
$\binom{M}{n'}$ spin states with excitation $n'$.
A typical  Hamiltonian in this scenario would be the $M$ atoms Tavis-Cummings Hamiltonian
$\sum_{\ell=1}^M\tfrac{\delta_\ell}{2}\sigma_\ell^z
+g_\ell(a^\dagger\sigma_\ell^-+a\sigma_\ell^+)$, and one could also think of interaction
between the spins like in the previous example as its contribution to the Hamiltonian
also commutes with the $I$ for this case. Plausible Lindblad operators are 
$\{a\}\cup\{\sigma_\ell^-\}_{\ell=1}^M$,
whereas for dephasing operators one could take $\{\sigma_\ell^z\}_{\ell=1}^M$.

\section{Conclusions}
We have presented a systematic method to solve  a broad class
of master equations in Lindblad form. 
Our approach is an alternative, but also extension to previous 
work \cite{Briegel1993,Barnett2000}. 
If the Liouville operator presents exclusively loss Lindblad
operators, we obtained the remarkable result 
that the full set of eigenvalues of the
Liouville operator is given by a sum of eigenvalues of
the effective non-Hermitian Hamiltonian used in the quantum trajectories approach.
When dephasing Lindblad operators are also present, 
the eigensystem can be obtained by diagonalizing
blocks of finite size of the full Liouvillian. 
We presented two examples of systems that can be solved in an analytical way as 
the solution involves the diagonalization of matrices of dimension of at most four.
For blocks larger than $4\times 4$, this approach can be used to improve
the numerical diagonalization, as one has to diagonalize blocks of smaller size than the dimension of 
the whole Hilbert space.

We also show as an application the analytical evaluation of the spontaneous emission spectrum of
an excited atom in an empty cavity. This shows that the introduced method can be manageable to
deal with analytical calculations. In this sense, it also offers the possibility to deal
with other more general class of Liouvillians in a perturbative manner. In particular
the extension to non-zero temperature baths comes to our mind as the Lindbladian term 
that accounts for the incoherent pumping or gain acts typically with a smaller strength.
In this way the eigensystem obtained with this construction might be used as a starting point to
solve more general problems using perturbation theory such as the evaluation
resonance fluorescence spectra \cite{Bienert2007,Torres2011} or the derivation of
reduced master equations \cite{Cirac1992}.

To finalize we would like to mention that these ideas could be also helpful in solving
systems like the coupled oscillators that describe  optomechanical systems \cite{Wilson-Rae2008}.
 In this case 
the free Hamiltonian of the optical mode serves also as a constant of motion and the
Lindblad operators of the mechanical mode are of the dephasing type.  

\begin{acknowledgements}
  The author would like to thank Toma\v{z} Prosen, Marc Bienert, Thomas H. Seligman 
and J. Zsolt Bern\'ad for enriching discussion and  useful comments. 
The postdoctoral grant CONACyT-162781 and the hospitality
of the Instituto de Ciencias F\'isicas, UNAM with the
project PAPIIT-IG101113 are also acknowledged.
\end{acknowledgements}


\end{document}